\documentclass[aps,prb,twocolumn,groupedaddress,showpacs]{revtex4}
\usepackage{epsfig}

\begin{document}


\title{Ultrafast quasiparticle relaxation dynamics 
in normal metals and heavy fermion materials}

\author{
K. H. Ahn, M. J. Graf, S. A. Trugman,
J. Demsar,~\cite{Demsar} 
R. D. Averitt, J. L. Sarrao, and A. J. Taylor 
} 
\affiliation{
Los Alamos National Laboratory, 
Los Alamos, New Mexico 87545}


\begin{abstract}
We present a detailed theoretical study of the ultrafast quasiparticle 
relaxation dynamics observed 
in normal metals and heavy fermion materials
with femtosecond time-resolved optical pump-probe spectroscopy.
For normal metals, a nonthermal electron distribution 
gives rise to a temperature ($T$) independent electron-phonon relaxation time
at low temperatures,
in contrast to the $T^{-3}$-divergent behavior predicted by the two-temperature model.
For heavy fermion compounds, we find that the blocking of electron-phonon
scattering for heavy electrons within the density-of-states peak near the Fermi energy 
is crucial to explain the rapid 
increase of the electron-phonon relaxation time below the Kondo temperature. 
We propose the hypothesis that the slower Fermi velocity compared to the sound velocity
provides a natural blocking mechanism due to energy and momentum conservation laws.
\end{abstract}


\maketitle

\newpage

\section{Introduction}

Recently, ultrafast time-resolved optical spectroscopy has been 
used to reveal the nature of the quasiparticle relaxation dynamics 
in condensed matter systems. 
Femtosecond time-resolved pump-probe optical measurements have been 
carried out in normal metals,~\cite{Groeneveld92,Groeneveld95}
conventional~\cite{Brorson90} and 
high-T$_c$ superconductors,~\cite{Stevens97} 
and charge-density wave solids.~\cite{Demsar99}
Time-resolved terahertz spectroscopy has measured
picosecond time-scale transient conductivity 
in colossal magnetoresistance manganites.~\cite{Averitt01}
These experiments show directly in the time domain how
constituent degrees of freedom of materials
interact with each other,
which is important to understand the physics 
governing the ground state and the low-energy excited states 
of materials.
Understanding the fast dynamics of quasiparticles is also
crucial for technological applications of these materials, 
for example, in the design of very fast switching devices. 

We recently reported femtosecond time-resolved pump-probe optical 
measurements on LuAgCu$_4$ and YbAgCu$_4$.~\cite{Demsar03}
These two materials are isostructural with 
a negligible difference 
in lattice constants~\cite{Pagliuso97} (about 0.3 \%) and 
about 1 \% difference in atomic mass between Lu and Yb.
However, their electronic structures are strikingly different 
due to the $f$-level occupancies in Lu and Yb.
The closed-shell $4f$ levels in Lu have no significant interaction with 
Cu $3d$ and Ag $4d$ conduction electrons, 
and LuAgCu$_4$ is a normal metal.~\cite{LuAgCu4}
In contrast, the open-shell $4f$ levels in Yb, 
that is, the localized hole in the $4f^{13}$ configuration,
have a strong interaction with conduction electrons,
and make YbAgCu$_4$ a heavy-fermion material,
which is characterized by a large peak 
in the electron density of states (DOS)
at the Fermi energy ($E_F$) and, equivalently, 
a large Sommerfeld coefficient $\gamma$.~\cite{Sarrao99,YbAgCu4}
Our time resolved optical experiments show 
very different relaxation dynamics in these two materials.
LuAgCu$_4$ shows a relaxation time ($\tau $) versus temperature ($T$) behavior
similar to other normal metals, such as Ag and Au.~\cite{Groeneveld92,Groeneveld95} 
In particular, there is very little $T$ dependence in $\tau $ at low temperatures.
However, YbAgCu$_4$ shows approximately a 100-fold increase in $\tau $,
as $T$ is decreased from the Kondo temperature ($T_K$) down to 10 K.
Kondo temperature is about 100 K in YbAgCu$_4$ and typically characterizes 
the width of the large DOS near $E_F$.

In Ref.~\onlinecite{Demsar03}, 
we presented experimental data showing different relaxation dynamics in
LuAgCu$_4$ and YbAgCu$_4$
along with the main theoretical ideas and final results of the calculations based on 
coupled electron and phonon Boltzmann transport equations.
In this paper, we report on the details of the theoretical model and analysis,
and clarify the underlying physics. 
The theories that existed before our study are summarized in Sec.~II~A.
Sections~II~B and III discuss
our calculations of the relaxation dynamics in normal metals and heavy fermion materials,
respectively.
Section~IV summarizes our results.

\section{Normal metals}

\subsection{Two-temperature model, its inconsistency with experiments,
and the nonthermal electron model}

In ultrafast optical pump-probe spectroscopy, 
an ultrafast laser pulse initially excites
the electron system, and 
the probe pulse monitors the relaxation of the electron system
by measuring transient optical properties
with subpicosecond time-resolution.
Because the diffusion of heat out of the illuminated (probed) region
is much slower (tens to hundreds of nanoseconds) 
than the time scales of interest, 
the relaxation of the excited electrons is due to 
the thermalization among electrons and other degrees of freedom, such as phonons, 
within the illuminated areas.

The most commonly used model for the relaxation dynamics
of photoexcited electrons in metals
is called the two-temperature model (TTM),~\cite{Kaganov56}
which assumes much faster relaxation 
{\it within} each constituent system
(e.g., electron system, phonon system)
compared to the relaxation {\it between} these constituent systems.
In this approximation, the temperature of each system can be defined during relaxation, 
and the relaxation time $\tau$ between system 1 and 2 is determined by 
their specific heats, $C_1$ and $C_2$, and 
the energy transfer rate per temperature difference, $g(T)$:~\cite{TTM.eq}
\begin{equation}
\frac{1}{\tau}=g(T) \left( \frac{1}{C_1} + \frac{1}{C_2} \right). \label{eq:TTM}
\end{equation}
Kaganov, Lifshitz, and Tanatarov~\cite{Kaganov56}
calculated the energy transfer rate $g(T)$ 
between electrons (e) and phonons (p) in normal metals
by solving coupled Boltzmann transport equations
for electrons and phonons with thermal equilibrium distributions at different temperatures.
Their results predict
$\tau \sim T$ at $T > T_{Debye}/5$ and 
$\tau \sim T^{-3}$ at $T < T_{Debye}/5 $,
where $T_{Debye}$ is the Debye temperature. 

Groeneveld {\it et al.}~\cite{Groeneveld92,Groeneveld95} have 
measured the relaxation time $\tau$ in Au and Ag
as a function of temperature ($T=10-300$ K)
and laser-energy density ($U_{laser}=0.3-1.3$ J cm$^{-3}$)
using the femtosecond pump-probe technique.
Their experimental data was inconsistent 
with the TTM predictions for
$\tau$ versus $T$ and $\tau$ versus $U_{laser}$, 
which led Groeneveld {\it et al.} to consider the nonthermal electron model (NEM).
Using a simple estimate, they pointed out  that the relaxation time
within the electron system is comparable to the electron-phonon 
relaxation time.
Instead of assuming a thermal equilibrium distribution for the electrons,
they numerically solved the Boltzmann equation 
for {\it electrons} 
with electron-electron and electron-phonon scattering,  
starting from an initial nonthermal electron distribution 
created by a laser pulse. 
(Based on the fact that the phonon specific heat is much larger 
than the electron specific heat,
phonons were assumed to have a thermal distribution 
with a time independent $T$.)    
This simulation could explain experimental data down to about 50 K.
Although Groeneveld {\it et al.}'s work proposed the essential idea
of a {\it nonthermal electron system},
their analysis focused on temperatures above 50~K, 
and excluded the low-$T$ region where 
the most striking difference between the TTM ($\tau \sim T^{-3}$)
and the experimental data (almost $T$-independent $\tau$) occurs.
Therefore, we undertake a more detailed analysis,
particularly focused on the low-$T$ region,
to understand the difference in relaxation dynamics 
between thermal and nonthermal electrons in normal metals
as well as in heavy fermion metals.

\subsection{Relaxation dynamics between nonthermal electrons and phonons}

We consider the coupled Boltzmann equations
for both {\it electrons and phonons}, with electron-electron and electron-phonon
scattering included.~\cite{pp}
The Boltzmann equations with momentum indices
are transformed into the following equations with energy indices
for a model with isotropic Debye phonons
and electrons with an isotropic parabolic dispersion relation.
In terms of electron and phonon distributions at time $t$, 
$f_{\epsilon}(t)$ and $b_{\omega}(t)$,
and the electron and phonon DOS, $D_e(\epsilon)$ and $D_p(\omega)$,
where $\epsilon$ and $\omega$ represent the electron and phonon energies,  
the coupled Boltzmann equations are~\cite{AM,Landau,Ziman}
\begin{eqnarray}
 \frac{df_{\epsilon}}{dt} &=& \left[ \frac{df_{\epsilon}}{dt} \right]_{ep}
+ \left[ \frac{df_{\epsilon}}{dt} \right]_{ee},  \label{Boltzmann.e} \\
 \frac{db_{\omega}}{dt}  &=& \left[ \frac{db_{\omega}}{dt} \right]_{ep},
\label{Boltzmann.p}
\end{eqnarray}
with the collision integrals
\begin{eqnarray}
\left[ \frac{df_{\epsilon}}{dt} \right]_{ep} 
&=& \int d\omega K_{ep} 
\{ [f_{\epsilon + \omega} (1-f_{\epsilon}) (b_{\omega}+1) 
\nonumber \\
&-& f_{\epsilon} (1-f_{\epsilon+\omega}) b_{\omega} ]
D_{p}(\omega) D_{e}(\epsilon+\omega)  
\nonumber \\
&+&[f_{\epsilon - \omega} (1-f_{\epsilon}) b_{\omega}
- f_{\epsilon} (1-f_{\epsilon-\omega}) (b_{\omega}+1) ]
\nonumber \\
& & D_{p}(\omega) D_{e}(\epsilon-\omega) \}, \label{eq:Boltzmann.1}
\end{eqnarray}
\begin{eqnarray} 
\left[\frac{db_{\omega}}{dt}\right]_{ep}
&=&\int d\epsilon  K_{ep} 
[ -b_{\omega} f_{\epsilon} (1-f_{\epsilon+\omega}) +
\nonumber \\
& &(b_{\omega}+1)f_{\epsilon+\omega} (1-f_{\epsilon}) ]
D_{e}(\epsilon) D_{e}(\epsilon+\omega),    \label{eq:Boltzmann.2}
\end{eqnarray}
\begin{eqnarray}
\left[ \frac{df_{\epsilon}}{dt} \right]_{ee}
&=&\int d\epsilon ' d\epsilon ''
\frac{1}{2} K_{ee} 
[
-f_{\epsilon} f_{\epsilon '} (1-f_{\epsilon ''}) (1-f_{\epsilon+\epsilon '-\epsilon ''})
\nonumber \\
&+&(1-f_{\epsilon})(1-f_{\epsilon '}) f_{\epsilon ''} f_{\epsilon+\epsilon '-\epsilon ''}
]  \nonumber \\
& &D_{e}(\epsilon ')  D_{e}(\epsilon '') D_{e}(\epsilon +\epsilon ' - \epsilon '' ). 
\label{eq:Boltzmann.3}
\end{eqnarray}
The linear dependence
of the electron-phonon scattering rate on the acoustic phonon energy $\omega$ 
[see, e.g., Eq.~(26.42) in Ref.~\onlinecite{AM}]
has been canceled out by a $1/\omega$ factor
which originates from changing momentum indices to
energy indices,
so that $K_{ep}$ is an energy independent constant.
$K_{ep}$ and $K_{ee}$ include the squares of the scattering matrix elements
and all other numerical factors, and 
have the units of [energy/time].

\begin{figure}
\leavevmode
\epsfxsize5.5cm\epsfbox{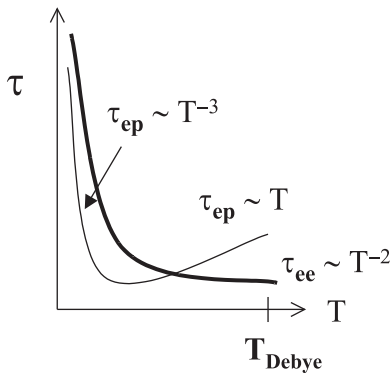}
\caption
{\label{fig:taueeep}
Schematic plot for 
the temperature dependence of  
electron-electron and electron-phonon relaxation times, 
$\tau_{ee}$ (thick line) and $\tau_{ep}$ (thin line).
$\tau_{ee}$ is from Eq.~(16) in Ref.~\onlinecite{Groeneveld95}.
$\tau_{ep}$ is from the TTM calculation, 
which assumes a {\it thermal} electron distribution.
The ratio between $\tau_{ee}$ and $\tau_{ep}$ 
at high $T$ (e.g., $T_{Debye}$ or 300 K) 
depends on the $K_{ee}/K_{ep}$ ratio.
For $K_{ee}/K_{ep}$=700 used in our calculation,
$\tau_{ee} < \tau_{ep}$ at $T=T_{Debye}$, as shown in the plot.
}
\end{figure}

In the remainder of this section,
we consider only light electron systems (normal metals).
The above Boltzmann equations, with electron and phonon DOS explicitly included,
show why the electron-electron relaxation time is comparable 
to the electron-phonon relaxation time at low temperatures.
The relaxation rate depends on the square of the scattering matrix elements
(i.e., $K_{ep}$ and $K_{ee}$)
and the number of available final states (i.e., $D_{e}$ and $D_{p}$).
If we compare Eqs.~(\ref{eq:Boltzmann.1}) and (\ref{eq:Boltzmann.3}),
even though  $K_{ee}$ is typically much larger than  $K_{ep}$,
the large disparity between the magnitude of the phonon ($D_p$) and electron ($D_e$) DOS
expedites the relaxation of electrons with phonons,
and hinders electrons from reaching thermal equilibrium.
The electron-electron and electron-phonon relaxation times,
$\tau_{ee}$ and $\tau_{ep}$, 
vary with $T$.
In Ref.~\onlinecite{Groeneveld95}, Groeneveld {\it et al.}  
found $\tau_{ee} \sim T^{-2}$ [Eq.~(16) of Ref.~\onlinecite{Groeneveld95}] 
from Fermi liquid theory.
Figure~\ref{fig:taueeep} schematically
compares the temperature dependence of $\tau_{ee}$ (thick line) and 
$\tau_{ep}$ (thin line), the latter of which is from the TTM results.~\cite{Kaganov56,Groeneveld95}
The plot shows that it is likely that
at high temperatures  
the electron system thermalizes within itself faster 
than the thermalization time with the lattice,
whereas at low temperatures 
electron-electron thermalization
becomes slower than the electron-phonon relaxation. 
More detailed discussion on $T$-dependent 
thermal versus nonthermal electron distributions 
will be given later in this section.

We now discuss the methods and parameter values 
we used to analyze the above Boltzmann equations.
The coupled Boltzmann equations are solved in two different ways.
In the first method, an initial condition at $t=0$ is set as
a nonthermal electron distribution 
excited by the laser pulse, $f_{\epsilon} (t=0)=f_{FD}(\epsilon,T_i) + \Delta f_{\epsilon}$,
and a thermal phonon distribution,
$b_{\omega}(t=0)=b_{BE}(\omega,T_i)$, 
where $T_i$,  $f_{FD}(\epsilon,T_i)$, and $b_{BE}(\omega,T_i)$
represent the initial temperature, Fermi-Dirac, and Bose-Einstein
distribution functions before the photoexcitation.
We find $f_{\epsilon}(t)$ and $b_{\omega}(t)$ at subsequent times by solving the differential 
equations 
until both $f_{\epsilon}(t)$ and $b_{\omega}(t)$ approach 
the final thermal distributions at equal $T$'s.
A small $\Delta f_{\epsilon}$ has been chosen so that
the final $T$ ($T_f$) is less than 3 K higher than $T_i$
(even at the lowest $T$),
which simulates the low laser intensity used 
in the experiments.~\cite{Demsar03}
In the simulation
the total electron energy $E_{e}(t)$
is calculated
at each time step, 
and the instantaneous relaxation time $\tau (t)$
is evaluated using 
$E_{e}(t)$ at three consecutive time steps.~\cite{3points} 

In the second method, we perform a linear stability analysis around the final states,
$f_{FD}(\epsilon,T_f)$ and $b_{BE}(\omega,T_f)$,
by expanding 
the coupled Boltzmann equations, Eqs.~(\ref{Boltzmann.e}) and (\ref{Boltzmann.p}),
linearly in $\delta f_{\epsilon}(t)$ and $\delta b_{\omega}(t)$, where
$f_{\epsilon}(t)=f_{FD}(\epsilon,T_f)+\delta f_{\epsilon}(t)$ 
and $b_{\omega}(t)=b_{BE}(\omega,T_f)+\delta b_{\omega}(t)$.
The zeroth order terms of the expansion vanish on both sides of the equations,
because they correspond to the final equilibrium state.
With
$\delta f_{\epsilon}(t)=v_{\epsilon}^e e^{-t/\tau}$, 
$\delta b_{\omega}(t)=v_{\omega}^p e^{-t/\tau}$,
and discretized $N$ electron and $M$ phonon energy levels,
the linear differential equations can be cast into 
an eigenvalue problem of an $(N+M) \times (N+M)$ matrix,
the solution of which gives $-1/\tau$ (eigenvalues) and 
the normal modes of the relaxation. 
Two of the normal modes have an unphysical infinite relaxation time $\tau$,
which originates from
total energy and total electron number conservation.
The rest of the modes represent all possible relaxation modes of the system.
In general, the relaxation of the system
can be represented as a linear combination of these modes.
By examine the eigenvector of each mode, 
we find whether the mode predominantly contributes to
$e$-$e$, $p$-$p$, or $e$-$p$ relaxation.
Since the energy transfer rate  
from the electron system to the phonon system 
at time $t$ 
of a specific mode with relaxation time $\tau$ is given by 
$dE_{e,\tau}(t)/dt=[\int (\epsilon -E_F) (-1/\tau) v^e_{\epsilon} D_e(\epsilon) d \epsilon ] e^{-t/\tau}$,
a useful quantity to identify the $e$-$p$ relaxation modes 
is the following $e$-$p$ energy transfer strength $r_E(i)$:
\begin{equation}
r_E(i)=\frac{1}{\tau(i)} 
\sum^{N}_{n=1} (\epsilon_n - E_F) v^e_n(i) D_e(\epsilon_n) \Delta \epsilon, \label{rE}
\end{equation}
where $\Delta \epsilon$ is the energy step size, and 
$(v^e_1, v^e_2, ..., v^e_N, v^p_1, ..., v^p_M)_i $
is the normalized eigenvector for mode $i$ with proper overall sign.
$r_E(i)$ characterizes the effectiveness of mode $i$
for electron-phonon energy relaxation. 
The modes which have large $r_E$ dominate in $e$-$p$ relaxation,
and their eigenvectors describe how 
the relaxation between the electron and phonon systems occurs.
The modes with small $r_E(i)$ are either $p$-$p$ or $e$-$e$ relaxation modes,
and describe how the relaxation within each system happens.

The electron and phonon energy levels are discretized
with a step size of 2 meV, which we find
small enough that any smaller step size would not
change our results even at 10 K (the lowest $T$ of the calculations and experiments).
The energy window for the electrons has been chosen to be 
between -0.15 eV and 0.15 eV with $E_F=0$,
which is wide enough in comparison to the highest $T$ of the calculations ($\sim$ 300~K).
Since the typical band width of normal metals
is of the order of eV, 
the electron DOS for normal metals is assumed 
to be constant within this energy window.
Fitting the experimental electronic specific heat ($C_e$) data,
we obtain $D_e$ = 2.1 eV$^{-1}$ f.u.$^{-1}$ spin$^{-1}$
for LuAgCu$_4$,~\cite{LuAgCu4}
which has been used for all of the results presented in this section.
Phonons are modeled by the Debye phonon model.   
Since only longitudinal phonon modes couple with electrons 
in the isotropic electron and phonon model,~\cite{Ziman}
we use the longitudinal phonon DOS with
a Debye energy $\omega_D$ = 24 meV (Ref.~\onlinecite{Sarrao99})
and 6 atoms per unit cell, $D_p (\omega)=18 \omega^2 / \omega_D^3$,  
in the Boltzmann equations. 
For most of the calculations, we use $K_{ee}/K_{ep}=700$
(see Figs.~\ref{fig:tau.Lu}, \ref{fig:tau.Lu.s.rep0},
and related discussion for this choice of $K_{ee}/K_{ep}$ ratio), and
$K_{ep}$ has been chosen as $K_{ep}=0.93$ eV/ps
for normal metals by fitting experimental data with the results of the calculation
(see Fig.~\ref{fig:tau.Lu}). 

\begin{figure}
\leavevmode
\epsfxsize5.5cm\epsfbox{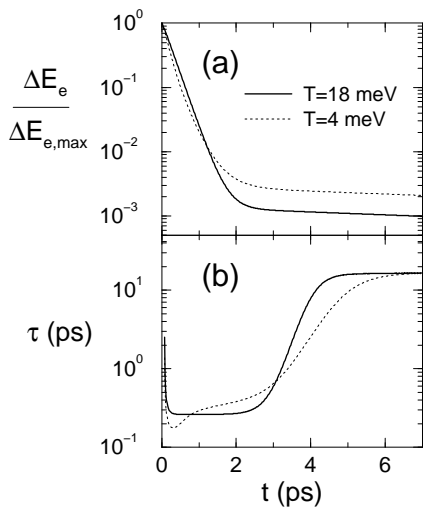}
\caption
{\label{fig:two.tau} 
A typical result of (a) normalized excess electron energy
versus time and 
(b) the instantaneous relaxation time versus time
at high $T$ (solid lines, $T$ = 18 meV = 208 K) and low $T$ (dotted lines, $T$ = 4 meV = 46 K).
}
\end{figure}

\begin{figure}
\leavevmode
\epsfxsize8.0cm\epsfbox{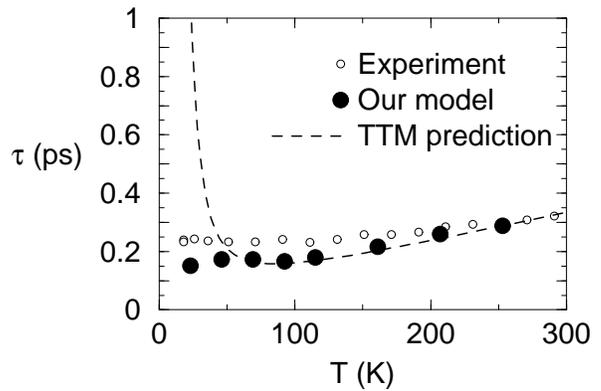}
\caption
{\label{fig:tau.Lu} 
Relaxation time $\tau$ 
calculated from the coupled Boltzmann equations
for LuAgCu$_4$,
along with experimental data
and the TTM prediction.~\cite{Demsar03}
}
\end{figure}

We now discuss the results obtained from the first method. 
Figure~\ref{fig:two.tau} shows a typical result 
of normalized excess electron energy, 
$\Delta E_{e} / \Delta E_{e,max}$ versus time ($t$),
and $\tau$ (instantaneous relaxation time) versus $t$
at high (solid lines) and low (dotted lines) temperatures,
where
$\Delta E_{e} (t)=E_{e}(t)-E_{e}(t= \infty )$ and 
$\Delta E_{e,max} = \Delta E_{e} (t=0)$.
Two different stages of relaxation can be identified
for high $T$ results.
During the intermediate time between 0.3 ps and 2 ps
(as mentioned earlier, the timescale in picosecond
is obtained by using $K_{ee}/K_{ep}=700$ and $K_{ep}$ = 0.93 eV/ps),
the energy transfer from electrons to phonons is fast.
After more than 99 percent of the excess electron energy is transfered
to phonons, 
a slow relaxation (about 100 times slower than the intermediate
stage) appears.  
The changes of distribution functions at different time stages
show that the fast relaxation during the intermediate time 
corresponds to electron-phonon relaxation, 
whereas the later slow relaxation corresponds to 
phonon-phonon thermalization processes.~\cite{late.tau}
The relaxation behavior is independent of the initial conditions,
i.e., the form of $\Delta f_{\epsilon}$,
except at {\it very} early stages of the relaxation,
namely, $t < 0.3$ ps in Fig.~\ref{fig:two.tau}.
Since the changes in electron distribution during
the late stage are expected to be unobservable due to experimental noise,
we identify the intermediate stage $\tau$
[$t= 0.3 \sim 2$ ps for the solid line in Fig.~\ref{fig:two.tau}(b)]  
as the one measured experimentally at high $T$.
As $T$ is lowered in the simulation,
the well-defined flat intermediate time region in the $\tau$ versus $t$ plot
is replaced by a gradually changing $\tau$ 
with a minimum as shown in Fig.~\ref{fig:two.tau}(b).
This indicates multiple relaxation times,
a sign of a nonthermal electron distribution.
As done in the experiments, we find the best fitting single relaxation time
through a linear fit of $\log | dE_{e}/dt |$ versus $t$, 
which is close to the minimum $\tau$ 
of the dotted line in 
Fig.~\ref{fig:two.tau}(b).

\begin{figure}
\leavevmode
\epsfxsize8.5cm\epsfbox{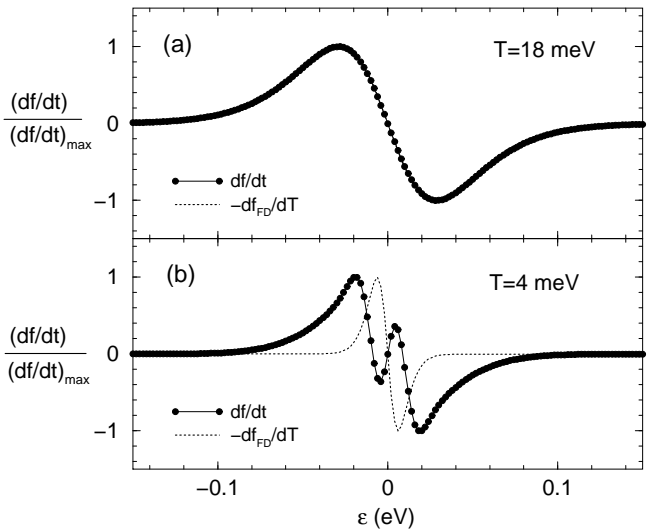}
\caption
{\label{fig:dfdt.Lu} 
Normalized $df/dt$ compared with 
normalized $-df_{FD}/dT$ at 
(a) high $T$  and 
(b) low $T$.
For high $T$, the two lines coincide.
}
\end{figure}

\begin{figure}
\leavevmode
\epsfxsize8.0cm\epsfbox{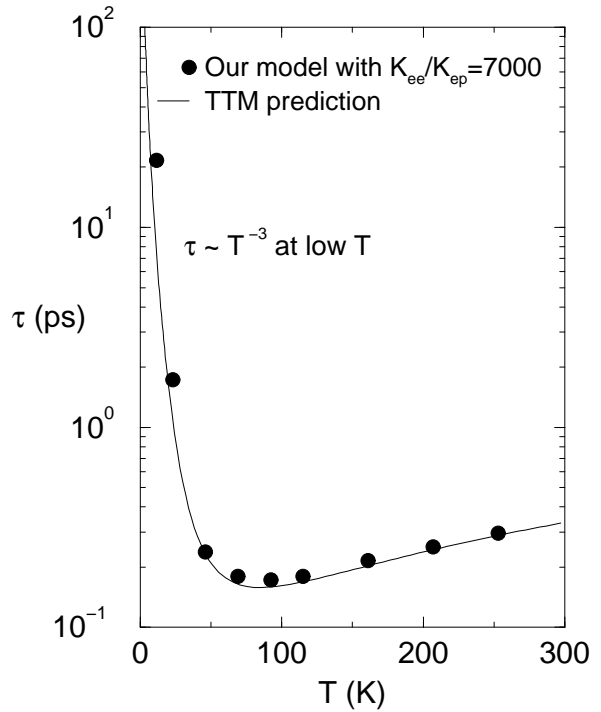}
\caption
{\label{fig:tau.Lu.s.rep0} 
Solid circles: calculated relaxation time $\tau$ from our model for LuAgCu$_4$
for a very large $K_{ee} \geq 7000 \times K_{ep}$.
Line: TTM prediction from Fig.~\ref{fig:tau.Lu}. 
With fast electron-electron relaxation 
our simulations based on coupled Boltzmann equations recover the TTM prediction 
$\tau \sim T^{-3}$ at low $T$.
}
\end{figure}

The $\tau$'s obtained at various $T$'s are
plotted in Fig.~\ref{fig:tau.Lu} (solid circles)
along with the experimental data (open circles) for LuAgCu$_4$
and the TTM prediction (dashed line, see Ref.~\onlinecite{Demsar03}).
The time unit has been scaled to fit the experimental data at high $T$,
which gives $K_{ep}$ = 0.93 eV/ps
as mentioned previously.
The fit
does not uniquely determine $K_{ee}$,
as long as $K_{ee}$ is smaller than or similar to $700 \times K_{ep}$.
The results show reasonable agreement with experimental data,
including the region below 50 K, where
the TTM predicts strikingly different 
$\tau \sim T^{-3}$ behavior (Refs.~\onlinecite{Groeneveld95} and \onlinecite{Kaganov56}).
By comparing $df/dt$ and $-df_{FD}/dT$,
we examine whether the electron system approaches 
the final equilibrium state
while maintaining a thermal distribution or not, as shown in 
Fig.~\ref{fig:dfdt.Lu} at high and low $T$'s. 
At high $T$ [Fig.~\ref{fig:dfdt.Lu}(a)], 
the two curves coincide with each other [indistinguishable in Fig.~\ref{fig:dfdt.Lu}(a)], 
implying that the electron system has a thermal distribution.
However, at low $T$ [Fig.~\ref{fig:dfdt.Lu}(b)], 
$df/dt$ (solid line with solid circles) has a width of the order 
of the Debye temperature ($T_{Debye}$)
instead of $T$, which is the width of $-df_{FD}/dT$ (dotted line).~\cite{ee}
The results clearly show that at low temperatures the electron system
does not have a thermal distribution, which agrees with the previous discussion
in relation to Fig.~\ref{fig:taueeep}.

The effect of a nonthermal electron distribution 
on the relaxation dynamics becomes even more evident, 
if we increase $K_{ee}$ to $7000 \times K_{ep}$
so that the electron-electron relaxation is faster than 
the electron-phonon relaxation even in the low-$T$ region.  
The result (solid circles) is shown in Fig.~\ref{fig:tau.Lu.s.rep0} 
along with TTM prediction (line).
In this case, the electron system has a thermal distribution in the whole $T$ range,
and the TTM prediction $\tau \sim T^{-3}$ is recovered at low $T$.
It is remarkable that our simulation with a large $K_{ee}$ recovers 
the TTM prediction
in spite of the fact that our Boltzmann equations do not include
direct phonon-phonon scattering and, therefore, the phonon distribution is 
nonthermal except at $t=\infty $.
We speculate that the very weak dependence 
of e-p thermalization dynamics 
on the phonon distribution
(in contrast to the very strong dependence
on electron distribution)
originates from the 
characteristics of Bose (versus Fermi)
statistics.~\cite{TTM.phonon}

\begin{figure}
\leavevmode
\epsfxsize6.0cm\epsfbox{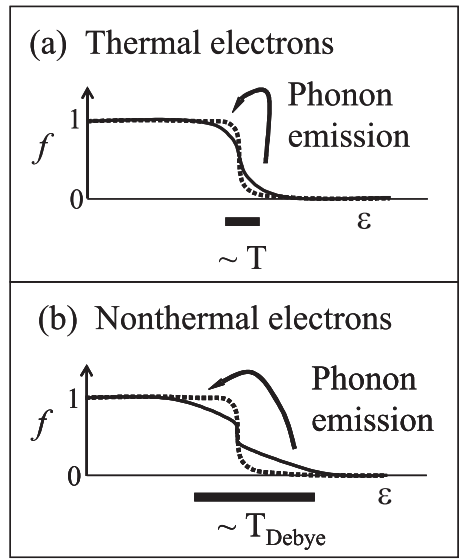}
\caption
{\label{fig:el.dist} 
Schematic pictures explaining different relaxation dynamics 
for (a) thermal and (b) non-thermal electron distributions at low temperatures.
Dotted lines represent the final thermal electron distribution.
}
\end{figure}

We provide a qualitative reason why the thermal electron
distribution gives rise to a slow $\tau \sim T^{-3}$ relaxation behavior
and the nonthermal electron distribution gives 
faster and less-$T$-dependent 
relaxation behavior at low $T$.
As depicted in Fig.~\ref{fig:el.dist}(a), 
if the electrons have a thermal 
distribution at $T$ (solid line), which is slightly higher than the final $T$ (dotted line),
the electron-phonon scattering important for 
the relaxation happens 
within the energy range of the order of $T$ from $E_F$.
Therefore, the relaxation rate depends on how many phonon modes
exist at $\omega < T$.
In the Debye phonon model, since $D_p(\omega) \sim \omega^2$,    
the relaxation rate $\tau^{-1}$ is proportional to $T^3$.
In contrast, 
if the electron distribution is nonthermal [solid line in Fig.~\ref{fig:el.dist}(b)],
then electron-phonon relaxation occurs in 
over an energy range of the order of Debye energy [see also Fig.~\ref{fig:dfdt.Lu}(b)].
This makes the electron-phonon relaxation faster and less $T$-dependent,
as indeed observed (Fig.~\ref{fig:tau.Lu}).

\begin{figure}
\leavevmode
\epsfxsize8.5cm\epsfbox{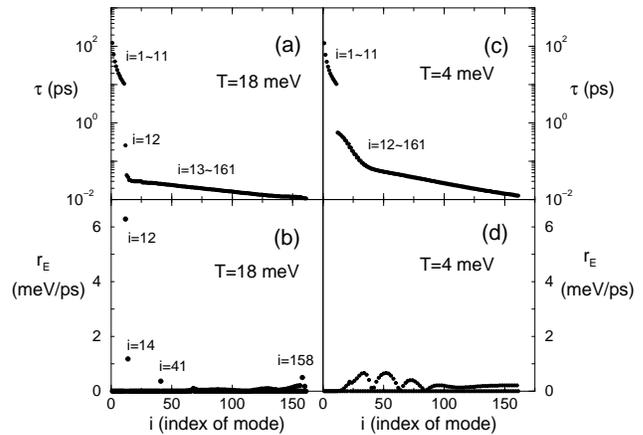}
\caption
{\label{fig:FPEA.Lu} 
[(a), (c)] Relaxation time $\tau$ and 
[(b), (d)] $e$-$p$ energy transfer strength $r_E$ 
at 
[(a), (b)] high $T$ and 
[(c), (d)] low $T$ 
calculated by the second method, 
which linearizes the coupled Boltzmann equations. 
The index $i$ is assigned in descending order of $\tau$.
}
\end{figure}

In the second method, 
we linearize the coupled Boltzmann equations
and calculate eigenvalues and eigenvectors of the matrix.
This second method supports and clarifies the results obtained 
earlier by the first method.
With the same discretized energy levels as in the first method,
a total of 163 energy levels (151 electron and 12 phonon energy levels) 
exist in our model, which means
a matrix of size $163 \times 163$ has to be numerically diagonalized.
Two of the 163 eigenmodes are unphysical as discussed earlier. 
The remaining 161 physical modes have negative eigenvalues, $-1/\tau$,
as expected for a stable fixed point, 
and represent 
all possible relaxation modes of the system.
The $\tau$'s at high and low $T$'s
are plotted in Figs.~\ref{fig:FPEA.Lu}(a) and \ref{fig:FPEA.Lu}(c), respectively.
The indices of modes $i$ are assigned in descending order of
$\tau$.  
The $e$-$p$ energy transfer strength $r_E$ calculated by Eq.~(\ref{rE})
for high and low $T$'s are plotted in 
Figs.~\ref{fig:FPEA.Lu}(b) and \ref{fig:FPEA.Lu}(d),
showing the efficiency of each mode for electron-phonon energy relaxation. 
At both $T$'s, there exist 11 modes 
(one less than the number of phonon energy levels) 
which have a much larger $\tau$ than the rest of the modes.
For example, at high $T$ these modes are 2 orders of
magnitude slower than the rest of the modes.
These slow modes, 
whose eigenvectors have mostly phonon components and very little electron components,
correspond to the late time phonon-phonon 
relaxation (see Fig.~\ref{fig:two.tau}).
For the high $T$ results in Figs.~\ref{fig:FPEA.Lu}(a) and \ref{fig:FPEA.Lu}(b),
a single mode with $i=12$ has a large $r_E$ and is very effective in
transferring energy from the electron system to the phonon system
in comparison to the other modes.
This mode has a value of $\tau$ 
identical to the intermediate time $\tau$ in Fig.~\ref{fig:two.tau} 
found by the first method,
and is well separated from other $\tau$ values as seen 
in Fig.~\ref{fig:FPEA.Lu}(a).~\cite{singular}
The electron part of its eigenvector matches $-df_{FD}/dT$,
confirming the results
obtained by the first method at high $T$ [Fig.~\ref{fig:dfdt.Lu}(a)],
that is, the presence of a thermal electron distribution.
Modes with $i$=14, 41, and 158 have smaller $r_E$
values
than the $i$=12 mode.
Their eigenvectors show that they participate in 
$e$-$p$ relaxation, but the electronic parts of 
the eigenvectors do not match $-df_{FD}/dT$.
Their nonthermal behavior seems suppressed in the 
experiments and in the simulation
due to the thermal $i$=12 mode, which has
more than 5 times larger energy transfer strength
$r_E$.
The remainder of modes between $i$=13 $\sim$ 161
with very small $r_E$ values are mainly
$e$-$e$ thermalization modes.
These features change as $T$ is lowered, as shown in  
Figs.~\ref{fig:FPEA.Lu}(c) and \ref{fig:FPEA.Lu}(d).
Many modes (modes with roughly $r_E >$ 0.1 meV/ps) 
transfer energy from the electrons to the phonons.
However, none of their corresponding electronic eigenvectors 
can be described by $-df_{FD}/dT$,
indicating the absence of a well defined electron temperature during 
relaxation.
This again supports the results of the first method,
presented in Fig.~\ref{fig:dfdt.Lu}(b).
The rest of the modes between $i$=13 $\sim$ 161
with $r_E$ very close to zero again correspond
to $e$-$e$ relaxation modes.
Importantly, these have larger $\tau$ values compared to
the high $T$ case,
showing the slowing down of 
$e$-$e$ relaxation at low $T$,
consistent with Fig.~\ref{fig:taueeep}.

Both methods show the importance of a nonthermal electron distribution 
for the electron-phonon relaxation dynamics at low $T$ for normal metals.
In the next section, we discuss how the relaxation dynamics 
of heavy fermion compounds are dramatically different from that
of normal metals. 

\section{Heavy fermion materials}

\begin{figure}
\leavevmode
\epsfxsize8.5cm\epsfbox{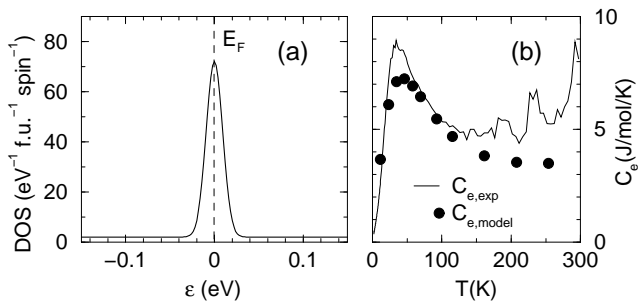}
\caption
{\label{fig:eDOS.Ce.Yb} 
(a) Model DOS and (b) calculated electron specific heat (solid circles) 
along with experimental data of $C_e$ (line) for YbAgCu$_4$.
}
\end{figure}

As mentioned in Sec.~I, heavy fermion materials
are characterized by a large DOS peak with a width of order $T_K$
near $E_F$. 
YbAgCu$_4$ has $T_K \sim$ 100 K 
and is a paramagnetic metal down to the 
lowest $T$ studied.~\cite{Sarrao99}
The interaction between localized $f$-electrons and
delocalized conduction electrons in heavy fermion materials
is a focus of intensive research
in strongly correlated electron systems
and has not yet been fully understood.
In this paper, we consider the simplest model
that captures the basic physics of heavy fermions, 
that is, a peak in the DOS near $E_F$.
Even though some theories~\cite{McQueen94}
predict a disappearance
of the peak DOS above $T_K$, we find that 
the calculated specific heat 
[Fig.~\ref{fig:eDOS.Ce.Yb}(b)] and 
relaxation time (Fig.~\ref{fig:tau.Yb}) have only 
a weak dependence on the presence or absence of the peak DOS
above $T_K$.
(Therefore, neither the relaxation time data nor the specific heat data
can distinguish whether the peak DOS is present or not above $T_K$.)
In the hybridization gap model,~\cite{Hewson} 
local $f$ levels hybridize with the conduction band
and open a gap with DOS peaks above and below the gap.
If $E_F$ is located within the peak, not in the gap,
the results we will show below have little dependence 
on whether we use a hybridization gap model or a single peak DOS model.
Therefore, in our simple model, 
we assume a $T$-independent single peak electron DOS
as shown in Fig.~\ref{fig:eDOS.Ce.Yb}(a).  
We further simplify the problem by choosing the $E_F$ at the center of the peak,
so that the chemical potential is $T$-independent.
The peak DOS is described by a Gaussian function
with a constant background DOS:
\begin{equation}
D_e(\epsilon)=D_{peak} \exp [-(\epsilon/\Delta)^2]
+D_0.  \label{eq:peakDOS}
\end{equation}
The discretized energy step size and energy window are identical
to the normal metal case in Sec.~II.
We determine the DOS parameters by fitting the electronic specific heat data, $C_e$,
as shown in Fig.~\ref{fig:eDOS.Ce.Yb}(b), and obtain
$D_{peak}=70$ eV$^{-1}$ f.u.$^{-1}$ spin$^{-1}$, 
$D_0=2.1$ eV$^{-1}$ f.u.$^{-1}$ spin$^{-1}$, 
and $\Delta = 13$ meV.
$D_{peak}$ is directly related to the linear slope of $C_e$ near $T=0$.
$\Delta$ is of the order of $T_K$ and determines the temperature 
of the peak in $C_e$.
$D_0$ is identical to 
the electron DOS at $E_F$ for LuAgCu$_4$,
and is related to $C_e$ above $T_K$.
Since LuAgCu$_4$ and YbAgCu$_4$ are isostructural with almost identical atomic masses,
we use identical phonon DOS for LuAgCu$_4$ and YbAgCu$_4$.

To gain insight into the nature of
the scattering between heavy electrons and phonons,
we consider a case in which only
heavy electrons with an isotropic parabolic dispersion
exist.
As in the normal metal case,
we transform the coupled Boltzmann equations with momentum indices
into the equations with energy indices. 
The result is similar to the normal metal case,
Eqs.~(\ref{eq:Boltzmann.1})-(\ref{eq:Boltzmann.3}),
except for one important difference:  
If the Fermi velocity $v_F$ is slower than the sound velocity $v_s $,
where $ v_F = ( \partial \epsilon / \partial k)_{\epsilon=E_F} $ 
and $ v_s = ( \partial \omega / \partial k)_{k \rightarrow 0} $,
then the phonon integration in the electron-phonon scattering has a 
lower bound, which represents 
the blocking of electron-phonon scattering for low energy phonons
and has a great influence on electron-phonon relaxation at low $T$.
Therefore, if $ v_F > v_s $, 
Eqs.~(\ref{eq:Boltzmann.1})-(\ref{eq:Boltzmann.3}) with a peak DOS for $D_e$
can be used to approximately model the relaxation dynamics for both light and heavy electrons, 
whereas if $ v_F < v_s $, appropriate blocking of scattering processes
should be imposed upon these equations.

\begin{figure}
\leavevmode
\epsfxsize8.0cm\epsfbox{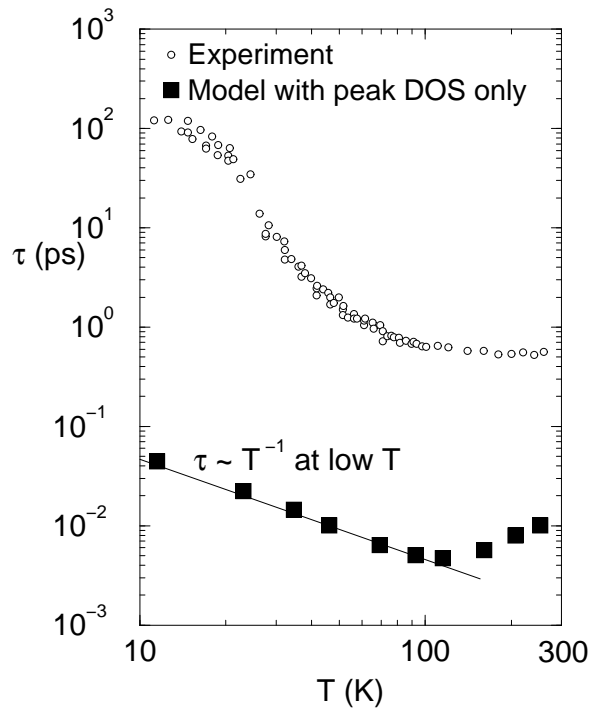}
\caption
{\label{fig:tau.Yb.wo.rst} 
Relaxation time calculated 
with the peak DOS only [Eq.~(\ref{eq:peakDOS}) and Fig.~\ref{fig:eDOS.Ce.Yb}(a)],
along with experimental data for YbAgCu$_4$. 
}
\end{figure}

We first consider the case $v_F > v_s$,
for which the important physics is simply
the increased DOS near $E_F$ given by
Eq.~(\ref{eq:peakDOS}).
We use the same value of $K_{ep}$ and $K_{ee}$ as the LuAgCu$_4$ case in Sec. II.
The calculated $\tau$ (solid squares) and  the experimental data (open circles)
shown in Fig.~\ref{fig:tau.Yb.wo.rst}
disagree in two respects.
First, the calculated $\tau$ at around $T$=300 K 
is about 60 times less than the observed $\tau$.
(Note that, experimentally, both YbAgCu$_4$ and LuAgCu$_4$ have similar $\tau$ 
at $\sim $ 300 K.)
This difference is due to the increased electron DOS in the calculation,
which enhances the e-p relaxation.
An approximately 60 times smaller $K_{ep}$ 
would shift the calculated $\tau$ in the whole $T$ range by 60 times, 
but it is unlikely that $K_{ep}$ in YbAgCu$_4$
would be smaller than $K_{ep}$ in LuAgCu$_4$
in such a drastic way.
Secondly, and more importantly, 
the divergence of calculated $\tau$'s at low $T$ is much weaker than the experimental data.
This slow divergence in calculated $\tau$ can be understood in the following way.
The large electron DOS increases both the electron-electron and the electron-phonon
scattering rates due to the increased number of available final states 
[see Eqs.~(\ref{eq:Boltzmann.1})-(\ref{eq:Boltzmann.3})].
However, electron-phonon scattering in Eq.~(\ref{eq:Boltzmann.1})
has only one factor of $D_e$, 
whereas electron-electron scattering in Eq.~(\ref{eq:Boltzmann.3})
has three factors of $D_e$.
Therefore, the electron-electron scattering rate increases much faster than 
the electron-phonon scattering rate, which makes the electron system thermal and
the TTM description valid.~\cite{OTM}
According to Kaganov {\it et al.}'s calculation,~\cite{Kaganov56} 
$g(T)$ in Eq.~(\ref{eq:TTM})
goes as $g(T) \sim T^4$ at low $T$ and $g(T) \sim constant$ at high $T$.
For normal metals, since $C_e \ll C_p$ even down to 10 K,
the TTM relation Eq.~(\ref{eq:TTM}) gives
$\tau \sim T^{-3}$.
In the case of heavy fermion compounds, 
$C_e$ is comparable to or larger than $C_p$ at low $T$,
and therefore $C_p \sim T^3$ for the Debye phonons 
produces $\tau^{-1} \sim T$ as the leading $T$ dependence.
Indeed, the low-$T$ divergence of the calculated $\tau$ in Fig.~\ref{fig:tau.Yb.wo.rst}
can be well described as $T^{-1}$. 
Therefore, the discrepancy with experimental data
cannot be fixed by simply changing parameters,
and it indicates that there is additional physics involved 
other than the large DOS at the Fermi level to explain
a 100-fold increase of $\tau$ below $T_K$.

If we take the approximation that
the difference between $E_F$ and the bottom of the heavy electron band 
is about $T_K$, then from the isotropic parabolic heavy electron dispersion model
and the carrier density of $n$ = 0.85 measured 
by the Hall effect for YbAgCu$_4$,~\cite{Sarrao99}
we can estimate $v_F \approx $  4 km/sec.
Ultrasonic measurements show that 
the longitudinal sound velocity along the [111] direction is
about 4.4 km/sec.~\cite{Zherlitsyn99}
This comparison shows that in fact
the two velocities are comparable to each other,
supporting the possibility of $v_F < v_s$.
We should note that experiments like de-Haas van-Alphen
are required for
a more reliable estimate of $v_F$.
Therefore, we hypothesize that 
the Fermi velocity is lower than the sound velocity,
and discuss how this affects the electron-phonon relaxation dynamics.

\begin{figure}
\leavevmode
\epsfxsize5.0cm\epsfbox{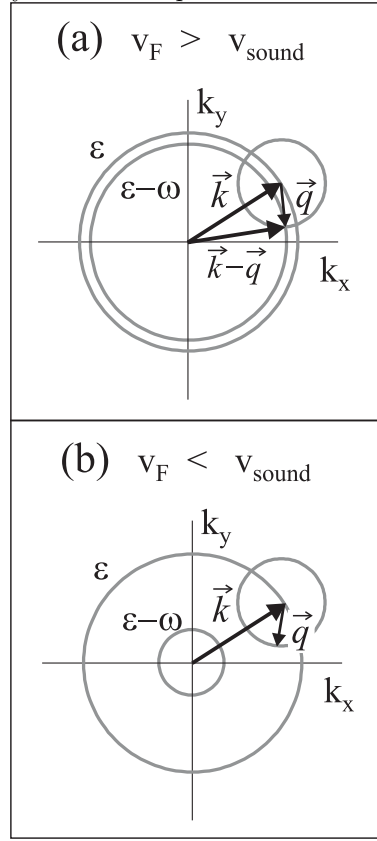}
\caption
{\label{fig:vfvs} 
Schematic pictures explaining different electron-phonon scattering processes 
for (a) $v_F > v_s$ and (b) $v_F < v_s$ cases.
If $v_F < v_s$, then the electron dispersion changes
more slowly than the phonon dispersion.
Therefore, the distance between 
the two $k$-space spheres with energies $\epsilon $ and $\epsilon - \omega $
is larger than the phonon momentum $q$, 
and no momentum and energy conserving scattering process is possible,
as shown in (b) in this figure.
}
\end{figure}

\begin{figure}
\leavevmode
\epsfxsize8.0cm\epsfbox{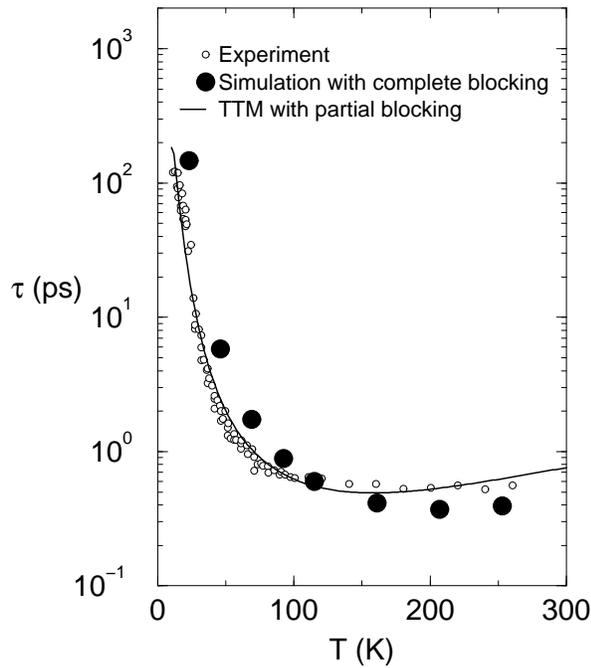}
\caption
{\label{fig:tau.Yb} 
Solid circles: relaxation time calculated with peaked DOS 
and blocking of e-p scattering within the peak DOS due to $v_F < v_s$.
Open circles: experimental data for YbAgCu$_4$.
Solid line: results obtained in Ref.~\onlinecite{Demsar03} using TTM 
with partial blocking of the e-p scattering within the peak.
}
\end{figure}

As depicted in Fig.~\ref{fig:vfvs}, 
if the Fermi velocity is slower than the sound velocity [Fig.~\ref{fig:vfvs}(b)],
then momentum and energy conservation requirements 
prohibit scattering between heavy electrons
and phonons.
In the actual heavy fermion DOS, heavy electron dispersion exists 
only within the peak DOS around $E_F$,
and electrons outside the peak
have a regular light mass.
Therefore, the above blocking mechanism 
applies only when both the initial and final 
electron states are within the peak of the DOS.
If one or both of the states are outside the peak,
regular electron-phonon scattering is expected.
These effects can be simply incorporated into the simulation
by blocking electron-phonon scattering within the peak DOS.
The results of the simulation are shown in Fig.~\ref{fig:tau.Yb}
as solid circles,
for which the blocked energy range is from -24 meV to 24 meV,
$K_{ep}$ = 0.23 eV/ps, and $K_{ee}=700 \times K_{ep}$.
This $K_{ep}$ value is smaller than the value for LuAgCu$_4$ case,
by about half in terms of the scattering matrix element.
It shows reasonable agreement with experimental data (open circles).

Since the electron system has a thermal distribution,
the TTM approach, with a similar blocking of electron-phonon scattering within the peak,
 can be used to describe the relaxation between 
the electrons and phonons,
which was discussed in Ref.~\onlinecite{Demsar03}.
In this TTM approach, we also included the possibility that some electron-phonon
scattering within the peak is allowed due to the anisotropy of  
$v_F(\hat{k})$ and $v_s(\hat{k})$, 
that is, $v_F > v_s$ along some directions and $v_F < v_s$ along other directions.
The results were shown as a solid line in Fig.~3 in Ref.~\onlinecite{Demsar03}
and reproduced in Fig.~\ref{fig:tau.Yb} (solid line),
which shows good agreement with experimental data.~\cite{light.HF}

The physical idea behind the increase of $\tau$ below $T_K$ is simple:
As $T$ decreases below $T_K$, or equivalently the width of the DOS peak,
since the electron system has a thermal distribution,
the main electron-phonon relaxation processes should occur 
within the peak [see Fig.~\ref{fig:el.dist}(a)], 
where the momentum and energy conservation laws block
the electron-phonon scattering.
Therefore, the electron-phonon relaxation time increases very rapidly 
as $T$ is lowered below $T_K$.~\cite{T.dep.pk}  
The results show that complete or substantial 
blocking of electron-phonon scattering processes within the DOS peak
is essential to explain the rapid increase of the electron-phonon 
relaxation time below $T_K$ in YbAgCu$_4$.

\section{Summary}

We provide a theoretical analysis of the ultrafast relaxation dynamics
observed by femtosecond time-resolved optical spectroscopy
in isostructural LuAgCu$_4$ and YbAgCu$_4$; the former is a normal metal
and the latter is a heavy fermion compound.
For normal metals, we find that a nonthermal electron distribution 
is responsible for 
a temperature-independent electron-phonon relaxation time
at low temperatures,
instead of a $T^{-3}$ divergent behavior predicted by the two-temperature model.
For heavy fermion compounds, we find that prohibiting electron-phonon
scattering within the density-of-states peak near the Fermi energy 
is crucial to explain the rapid 
increase of the electron-phonon relaxation time below the Kondo temperature. 
On the basis of the estimated Fermi velocity and the measured sound velocity,
we propose the hypothesis that the slower Fermi velocity compared to the sound velocity
provides this blocking mechanism due to energy and momentum conservation laws.
We find good agreement between the experimental data and our model
for both normal metals and heavy fermion compounds.

We thank V. V. Kabanov for useful discussions. 
This work has been supported by U.S. DOE.


\begin{thebibliography}{99}
\bibitem{LuAgCu4}
LuAgCu$_4$ has 
a 15 times larger density of states at the Fermi energy ($E_F$) 
compared to Ag
due to the large number of bands at $E_F$, 
not due to the hybridization of
local $f$-levels and conduction levels.
\bibitem{YbAgCu4}
The Sommerfeld coefficient $\gamma$ for YbAgCu$_4$ is 210 mJ/mol K$^2$,~\cite{Sarrao99}
which corresponds to a 300 times larger DOS at $E_F$ compared to Ag.
\bibitem{TTM.eq}
The TTM result is found from
$ C_1 \frac{dT_1}{dt} =- C_2 \frac{dT_2}{dt} = -g(T) (T_1 - T_2) $.
\bibitem{pp}
We have examined the effects of phonon-phonon scattering,
and found that it has
a negligible effect on the electron-phonon relaxation time.
\bibitem{3points}
From the relation $ \tau = ( - dE_e/dt ) /  ( d^2E_e/dt^2)$ for
$E_e(t)=E_{0}+\Delta E \exp (-t/\tau)$, 
we use  
$ \tau (t) =  - [E_e(t+dt)-E_e(t-dt)] dt / \{ 2 [E_e(t+dt)-2 E_e(t)+ E_e(t-dt)] \}$.
\bibitem{late.tau}
The phonon-phonon thermalization process occurs via electron-phonon scattering,
since our Boltzmann equations do not include direct phonon-phonon scattering. 
For example, a high energy phonon is absorbed by an electron,
which emits two low energy phonons.
\bibitem{ee}
The presence of the extra feature near $E_F$ 
depends on details of the parameter values and time at which $df/dt$ is evaluated.
When this feature is present, as shown in Fig.~\ref{fig:dfdt.Lu}(b),
it is partly due to electrons heating up
near $E_F$ through electron-electron scattering.
(Since electron-electron scattering conserves 
total electron energy, 
$(df/dt)_{ee}$ must cross zero at least three times.)
\bibitem{singular}
When the step size for energy discretization 
is reduced by half, 
the gaps in $\tau$ between this single mode and the edges
of the continuum-like faster and slower $\tau$'s
do not change.
This singular behavior deserves future theoretical studies.
\bibitem{OTM}
We find that the electron thermalization recovers the TTM predictions,
as in normal metals (see Fig.~\ref{fig:tau.Lu.s.rep0}).
\bibitem{light.HF}
The hypothesis that the relation $v_F < v_s$ provides
the blocking mechanism indicates that
the heavy fermion metals with lighter effecttive masses
may show relaxation dynamics 
qualitatively different from that of YbAgCu$_4$.
\bibitem{T.dep.pk}
Even when the electron DOS is $T$-dependent,
as long as the peak width in the DOS does not
increase faster than $T$,
the results do not change much.
\end{thebibliography}
\end{document}